\documentclass[twocolumn]{revtex4-2}
\usepackage{hyperref}
\usepackage{amsmath, amsthm}
\usepackage{graphicx}

\theoremstyle{definition}
\newtheorem{example}{Example}
\newtheorem{remark}{Remark}

\begin{document}
\title{An Improved Algorithm for Coarse-Graining Cellular Automata}
\author{Yerim Song}
\affiliation{Canyon Crest Academy, San Diego, CA, USA}
\email{yerim1111@gmail.com}
\author{Joshua A. Grochow}
\affiliation{Department of Computer Science, University of Colorado Boulder, Boulder, CO, USA}
\affiliation{Department of Mathematics, University of Colorado Boulder, Boulder, CO, USA}
\email{jgrochow@colorado.edu}

\begin{abstract}
In studying the predictability of emergent phenomena in complex systems, Israeli \& Goldenfeld (\emph{Phys. Rev. Lett.}, 2004; \emph{Phys. Rev. E}, 2006) showed how to coarse-grain (elementary) cellular automata (CA). Their algorithm for finding coarse-grainings of supercell size $N$ took doubly-exponential $2^{2^N}$-time, and thus only allowed them to explore supercell sizes $N \leq 4$. Here we introduce a new, more efficient algorithm for finding coarse-grainings between any two given CA that allows us to systematically explore all elementary CA with supercell sizes up to $N=7$, and to explore individual examples of even larger supercell size. Our algorithm is based on a backtracking search, similar to the DPLL algorithm with unit propagation for the $\mathsf{NP}$-complete problem of Boolean Satisfiability.
\end{abstract}

\maketitle

\section{Introduction}
Cellular automata (CA) are a model of dynamical systems that are discrete in both space and time. Since their development by Ulam and von Neumann \cite{vonNeumann} in the 1940s, CA have been applied to many subjects including biology \cite{bio}, physics \cite{ilachinski}, and computer science \cite{mitchell, sarkar}. 

Coarse-graining is one method of reducing the complexity of a system, while hopefully retaining enough structure to reveal insights about it. Rather than focusing on small, specific details, we ``zoom out'' and look at a larger picture. Israeli \& Goldenfeld \cite{IG1, IG2} introduced a method of coarse-graining cellular automata, and applied it systematically to the \emph{elementary} cellular automata. 

Elementary cellular automata were introduced and studied extensively by Wolfram [6]: they are automata with 2 states, on a 1-dimensional grid of cells, with only nearest-neighbor interactions. There are exactly 256 distinct elementary CAs, and Wolfram proposed dividing them into four classes based on their long-term behavior: stabilizing to a homogeneous state, ultimately becoming periodic, or maintaining random or complex-looking behavior indefinitely. The latter classes are hypothesized to be ``computationally irreducible'' and difficult or impossible to predict \cite{wolframBook, wolframPhysica, wolframNature}. However through the process of coarse-graining, even CA that are originally in these complex classes can coarse-grain to CA in the simpler classes \cite{IG1, IG2}.

To calculate these coarse-grainings, Israeli and Goldenfeld used a brute force algorithm, which required $2^{2^N}$ steps to search for coarse-grainings of supercell size $N$. The lengthy time required for the computations limited their study to coarse-grainings of supercell size only 4. They reported that there were 16 elementary CA that they could not coarse-grain at all, and for the other CA it was unknown whether they had found all the available coarse-grainings.

In this paper we develop a new algorithm, using a backtracking search similar to the DPLL algorithm for Boolean Satisfiability \cite{DP, DLL}. This backtracking search allows us to prune branches from the search tree early, based on coarse-graining constraints. Furthermore, dynamically ordering the variables allows us to speed up the process considerably in practice. 

Using our new algorithm, we were able to systematically find all coarse-grainings between elementary CA up to supercell size 7 on a commodity laptop. Note that $2^{2^7}\approx 3.4 \times 10^{38}$, so even if each step could be done in 1ms the brute force algorithm of \cite{IG1, IG2} would take $\approx 10^{27}$ years to handle supercell size 7. We were also able to search for specific coarse-grainings of larger supercell size. We find 56 new coarse-grainings at these larger supercell sizes (26 up to symmetry), though we leave open the mathematical question of how to prove when all nontrivial coarse-grainings have been found. Interestingly, all of these new coarse-grainings were found at $N=5$ or $6$; there were no new pairs of CA $A,B$ such that $A$ coarse-grained to $B$ at supercell size $7$.

\subsection{Related Work}

Magiera \& Dzwinel \cite{MD14, DM15} also present an improved algorithm for coarse-graining CA, and get up to supercell size 7. However, despite the similar name, they are solving a related but different problem than the one we solve. Namely, they solve the problem:
\begin{quotation}
\noindent \textsc{IsCoarseGrainable}

\noindent \textit{Input:} A CA $A$ and supercell size $N$

\noindent \textit{Output:} A nontrivial coarse-graining of $A$ with supercell size $N$, or $\bot$ if none exists.
\end{quotation}
In contrast, we solve the following problem, which enables us to fill out Fig.~\ref{fig:diagram} (a CA analogue of a renormalization group flow diagram):
\begin{quotation}
\noindent \textsc{CoarseGraining}

\noindent \textit{Input:} Two CAs $A,B$ and supercell size $N$

\noindent \textit{Output:} A nontrivial coarse-graining from $A$ to $B$ with supercell size $N$, or $\bot$ if none exists.
\end{quotation}
Their algorithm begins by constructing $A^N$, with $k^N$ states per cell if $A$ had $k$ states per cell, and then tries to collapse those new states together as much as possible, resulting in another CA (but one which is not specified ahead of time in the input). In contrast, our algorithm tries to find the coarse-graining map from $A^N$ to the given CA $B$. The fact that $B$ is given in the input allows us to use other algorithmic tactics that are less easy to take advantage of in their setting. In contrast, their algorithm is aimed at deciding which CA can be coarse-grained at all.

We note that one could directly reduce to a Boolean Satisfiability problem and use an off-the-shelf SAT solver. This did not seem very promising to us, as supercell size $N$ would result in a SAT instance with $2^N$ variables and $2^{3N}$ constraints, where each constraint would consist of several CNF clauses. See Remark~\ref{rmk:SAT} below for more details.

However, given that our approach is similar to approaches to the $\mathsf{NP}$-complete problem of Boolean Satisfiability, it is natural to wonder about comparing the two. Because the coarse-graining problem has exponentially many variables---one variable for the projection of each possible $N$-tuple of states---it is easy to see that \textsc{CoarseGraining} is in the complexity class $\mathsf{NEXP}$. It would be interesting to know whether it is $\mathsf{NEXP}$-complete, as that would suggest that some amount of brute force search is inevitable, assuming the widely believed complexity conjecture that $\mathsf{EXP} \neq \mathsf{NEXP}$.

We note that while we present our algorithm and results in the context of \emph{elementary} CA (2-state, 1D, nearest neighbor), it is trivial to adapt it to arbitrary CA.

\section{Background on Cellular Automata and Coarse-Graining}
A cellular automaton (CA) $A$ has cells at the nodes of some (usually infinite, regular) graph, each cell has a state from a finite set $S_A$, and an update rule $f_A(a_0; a_1, \dotsc, a_k)$, which takes in the state $a_0$ of a cell and the states $a_1, \dotsc, a_k$ of its neighbors, and outputs the state of the cell for the next time step. In this paper we consider only \emph{elementary} CA, which are CA on a 1D lattice, with nearest-neighbor interactions, and 2 states, $S_A = \{0,1\}$. Thus each rule $f_A$ takes in only three states: that of a cell, its left neighbor, and its right neighbor. For geometric convenience, rather than putting the cell's state first, we put them in geometric order, so that, e.g., $f_A(a_1,a_2,a_3)$ is the next state of the cell in position 2. 

As introduced by Israeli \& Goldenfeld \cite{IG1, IG2}, a coarse-graining from one CA $A$ to another CA $B$ with supercell size $N$ is a map $P\colon S_A^N \to S_B$ such that, given any string states, if we first run $A$ for $N$ steps and then apply $P$, the result is the same as first applying $P$, and then running $B$ for a single step. That is, the coarse-graining scales both the cell size and the speed of the CA by $N$. 

For 1D CA with nearest-neighbor interactions, this is captured precisely by the fundamental coarse-graining equation
\begin{equation}
P(f_A^N(x_1, x_2, x_3)) = f_B(P(x_1), P(x_2), P(x_3)) \quad \forall x_i \in S_A^N \label{eq:CG}
\end{equation}
Given an input $(x_1, x_2, x_3) \in S_A^{3N}$, we will refer to the result on the left-hand side here throughout the paper as $res_1=res_1(x_1, x_2, x_3)$, and similarly the result on the right-hand side as $res_2$. We will sometimes abbreviate Eq.~(\ref{eq:CG}) to:
\[
PA^N = BP.
\]
Given CA $A, B$ and supercell size $N$, our goal is thus to find a projection $P:S_A^N \to S_B$ such that 
\[
res_1=res_2 \quad \forall (x_1, x_2, x_3) \in (S_A^N)^3.
\]
These will be the fundamental constraints that our backtracking algorithm explores and exploits.

Following \cite{IG1, IG2}, we consider a coarse-graining ``trivial'' if $P$ is a constant function, that is, if it maps all inputs to $0$ (resp., all inputs to $1$). We ignore these trivial coarse-grainings by fiat.

It is difficult to discover which projection functions $P$ are valid coarse-grainings at window size $N$, because there are $2^{2^N}-2$ possibilities for $P$ (excluding the two trivial possibilities), and for each such $P$ there are $2^{3N}$ constraints to check. There seems to be no apparent pattern as to which projections will satisfy all of constraints, and thus be a coarse-graining. (The lack of such patterns could be attested to theoretically by answering affirmatively our question above about $\mathsf{NEXP}$-completeness.) In the absence of such patterns, one might be resigned to trying every possible projection, and indeed this is essentially the approach taken in \cite{IG1, IG2}. 
In this paper we show that a backtracking search with a few easy-to-implement heuristics can do significantly better in practice, while still ensuring our algorithm is \emph{complete} in the sense that every possible coarse-graining will be found.

\subsection{Symmetries}
In our algorithm, we will determine whether there are coarse-grainings from $A$ to $B$ for all pairs of elementary CA $A,B$, for supercell sizes up to $N=7$. Once we have found a coarse-graining from $A$ to $B$, we do not consider it further in our experiments; e.\,g., if we find a coarse-graining from $A$ to $B$ at supercell size $3$, we do not search for coarse-grainings from $A$ to $B$ with larger supercell sizes, though our algorithm could be used to do so for further exploratory purposes.

Additionally, as noted in \cite{IG1, IG2}, we may eliminate CA that are ``isomorphic'' to one another under the symmetries that swap left and right, that swap 0 and 1, or that swap both left-right and 0-1. If we use $\sigma_{LR}$ to denote the left-right swap, $\sigma_{01}$ to denote the $0$-$1$ swap, and $A \to B$ to denote a coarse-graining, then we have:
\begin{align*}
A \to B \Longleftrightarrow & \sigma(A) \to \sigma(B) \\
& \forall \sigma \in \{\sigma_{LR}, \sigma_{01}, \sigma_{LR} \circ \sigma_{01} = \sigma_{01} \circ \sigma_{LR}\}.
\end{align*}

For example, suppose $A$ is rule 24 and $B$ is rule 240. We have $\sigma_{01}(24)=231, \sigma_{LR}(24)=66, \sigma_{01,LR}(24)=189$, and $\sigma_{01}(240)=240, \sigma_{LR}(240)=\sigma_{LR,01}(240)=170$. So among the possibilities
\[
24 \to 240 \qquad 231 \to 240 \qquad 66 \to 170 \qquad 189 \to 170,
\]
we need only check one, rather than all four. By eliminating these symmetric cases, the number of comparisons we must make between pairs of elementary CA is cut down to about 1/3 of the original total of $256^2$; it is not 1/4 because of the presence of some rules that are mapped to themselves by some of the symmetries (such as $\sigma_{01}(240)=240$ in the preceding example).

\section{Backtracking Algorithm}
The new approach we developed to more efficiently compute coarse-grainings involves considering a backtracking tree search, allowing us to to selectively cut tree branches earlier in the computation without losing information. The basic idea is that, given cellular automata $A,B$ and supercell size $N$, we seek to find values for $P(x)$ for all $x \in S_A^N$, that satisfy (\ref{eq:CG}). For each such $x$, we branch on trying either $P(x)=0$ or $P(x)=1$ (when working with elementary CA). As we assign values to $P(x)$ for some of the $x$'s, we can begin to see whether (\ref{eq:CG}) is satisfied or violated for different inputs $(x_1,x_2,x_3) \in S_A^{3N}$. In some cases, this lets us undo an assignment for $P(x)$, backtrack, and try the other value. 

\begin{figure}[!htbp]
\includegraphics[width=\columnwidth]{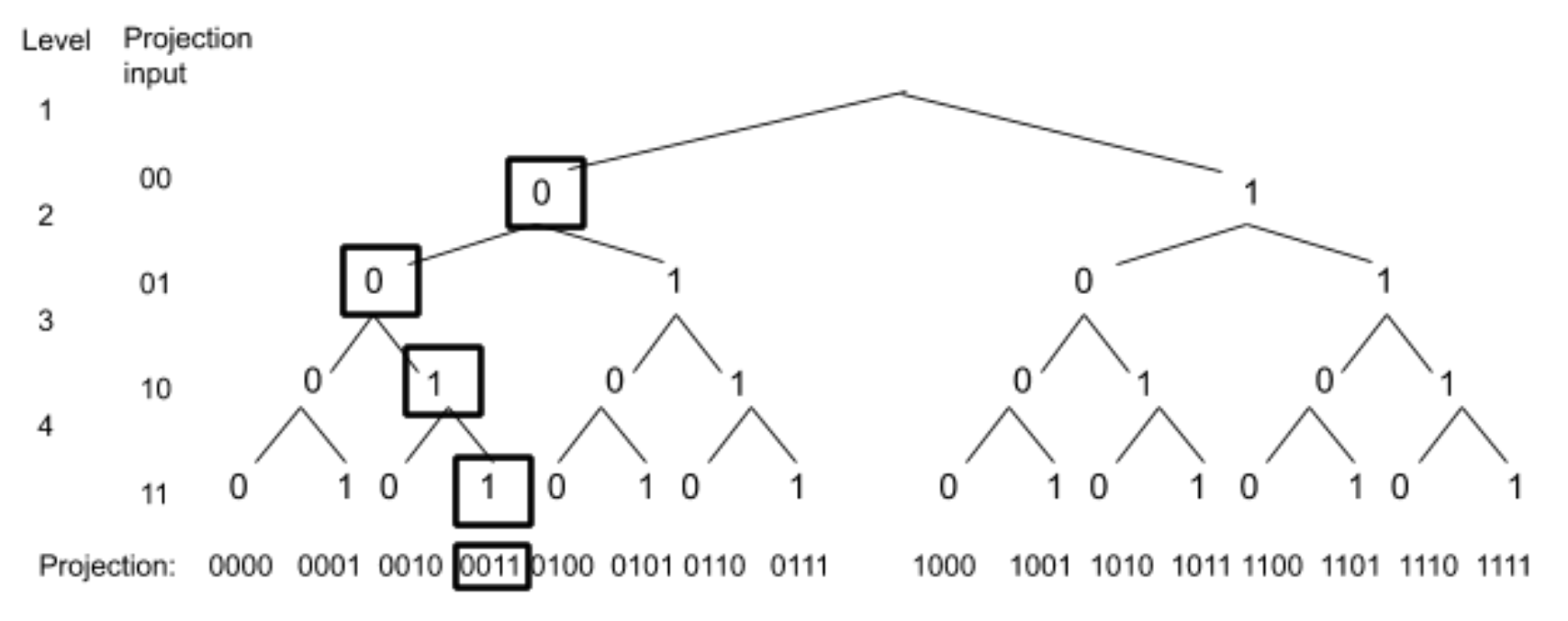}
\caption{\label{fig:tree} The search tree above displays all 16 possible projections with supercell size $N=2$. Based on the projection input on the left, the output is displayed on the tree. In this diagram, level 1 represents $P(00)$, level 2 represent $P(01)$, and so on. The boxed sequence demonstrates that the fourth leaf corresponds to the projection $0011$, that is, $P(00)=0, P(01)=0, P(10)=1, P(11)=1$.}
\end{figure}

Figure~\ref{fig:tree} shows an example of this search tree for supercell size $N=2$. 
Generally, for supercell size $N$, the tree will have $2^N$ levels and $2^{2^N}$ leaves, though many of these will be pruned before they are reached. Note that although it is included in the tree diagram, we exclude the all-0s and all-1s projections in our calculations, as these are considered ``trivial'' coarse-grainings, as in Israeli \& Goldenfeld. These projections are represented in the first and last leaves in the diagram. If we did not exclude them, then almost all rules could be coarse grained by these trivial projections.

\subsection{Tree pruning}
The first way we reduce the number of computations is to prune branches at earlier levels of the tree, so that computations in the remaining levels can be skipped. By allowing several projections to be skipped early on, we manage to significantly reduce the computation time to find or rule out a coarse-graining.

To determine whether a branch can be pruned the algorithms checks (\ref{eq:CG}) on inputs $(x_1, x_2, x_3) \in (S_A^N)^3$ that have been determined by the choices made so far. Given a partial projection $P$, defined only on a subset of $S_A^N$, we say that an input $(x_1, x_2, x_3)$ has a decided projection if $P(x_1), P(x_2), P(x_3)$, and $P(f_A^N(x_1, x_2, x_3))$ have all been assigned already. 
We begin at level 1, where the first projection will be set to either 0 or 1. Meanwhile, the rest of the projections will be temporarily undecided. Using all possible inputs with decided projections, we check if $PA^N=BP$ is satisfied before proceeding to the next level. Recall that we use $res_1$ to denote the output of $PA^N$ (when defined) and $res_2$ the output of $BP$. If we discover a decided input for which $res_1 \neq res_2$, then the most recent branch leading to the current projection can be safely cut off, as there are no valid projections that extend it (i.e., in its subtree).

\begin{figure}[!htbp]
\includegraphics[width=\columnwidth]{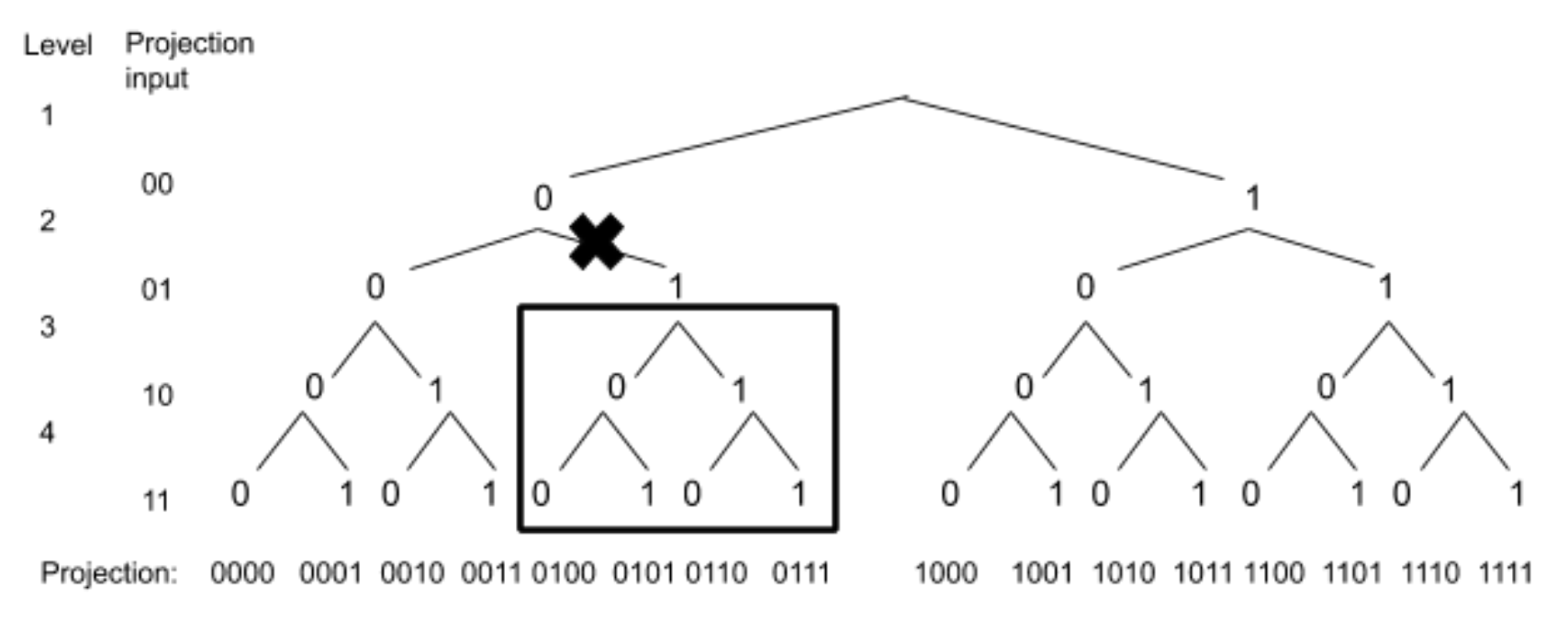}
\caption{\label{fig:cutbranch} Tree pruning. The search tree for attempting to coarse grain from rule 196 to rule 192 at supercell size $N=2$ (Example~\ref{ex:cutbranch}). After branching on $P(00)=0$ and $P(01)=1$, one of the equations (\ref{eq:CG}) is violated, and the search backs up to try the branch $P(01)=0$ instead. The `X' represents the tree being pruned at level 2; the box represents all the computations that were skipped due to this cut.}
\end{figure}

\begin{example} \label{ex:cutbranch} To better illustrate the method, we walk through an example of finding a coarse-graining from rule 196 to rule 192 with supercell size $N=2$; this example is illustrated in Figure~\ref{fig:cutbranch}. To begin, the tree diagram will start at level 1 and branch on $P(00)=0$. We only check inputs whose projections are defined, so the only available input to test at this point is $x=000000$. When rule 196 is applied to this pattern $N=2$ times, the result is $00$, so we may indeed check the result at $x$. $P(00)=0$, so $res_1=0$. 
To calculate $res_2$, we apply $P$ to the input $000000$ three times, resulting in $P(00)P(00)P(00)=000$, and then use rule 192 on $000$ to get $res_2=0$. Because $res_1=res_2$ at this point, we cannot cut this branch, and continue to the next level in the tree.

Suppose that in the next level we branch on $P(01)=1$ (and we still have assigned $P(00)=0$). As $x=000100$ has its projection defined, we may attempt to validate our choices on that input. To get $res_1$ we first apply rule 196 twice to get an outcome of $01$ and then $P$ to the result to get $PA^N (000100)=P(01)=1$. To compute $res_2$ we apply rule 192 to $P(00)P(01)P(00)=010$, resulting in $res_2=0$. Since $res_1 \neq res_2$, we can cut off the whole branch with $P(00)=0, P(01)=1$, avoiding further computations. 
\end{example}

\begin{remark} \label{rmk:SAT}
This is one of the places our approaches differs from simply reducing to a Boolean Satisfiability (SAT) problem and using an off-the-shelf SAT solver. In our approach, we get to choose ``on the fly'' which inputs $(x_1, x_2, x_3)$ to test for $res_1 = res_2$. In contrast, a reduction to a SAT solver would require writing down \emph{all} the constraints for every possible input $x$ right from the beginning (of which there are $8^N$), before any branches have been made. 
\end{remark}

\subsection{Forced values}
Our second method to reduce computation time is to leverage forced values of the projection. Namely, once we have branched on assigning $P(x)$ for some values of $x \in S_A^N$, we may be forced to assign $P(x')$ for some new, previously unassigned $x'$ in order to satisfy (\ref{eq:CG}). (This is similar to unit propagation in Boolean Satisfiability solvers.) When this occurs later in the tree, it lets us eliminate duplicated computations; when it occurs earlier in the tree, it lets us avoid large swaths of the tree. 

In this method, instead of simply checking whether $res_1 = res_2$ on inputs $(x_1, x_2, x_3)$ for which $P(x_1), P(x_2), P(x_3)$, \emph{and} $P(f_A^N(x_1, x_2, x_3))$ have all been assigned, we make an effort to match $res_1$ and $res_2$ even when these have not yet all been assigned. As in the method and example above, we begin with an input $(x_1, x_2, x_3)$ such that $P(x_1), P(x_2)$, and $P(x_3)$ are all defined. We can then calculate $res_2 = f_B(P(x_1), P(x_2), P(x_3))$. Let $x_4 = f_A^N(x_1, x_2, x_3)$; note that $res_1 = P(x_4)$. If $P(x_4)$ is not yet assigned, then the value of $P(x_4)$ is forced by our earlier choices to be equal to $res_2$. 

\begin{figure}[!htbp]
\includegraphics[width=\columnwidth]{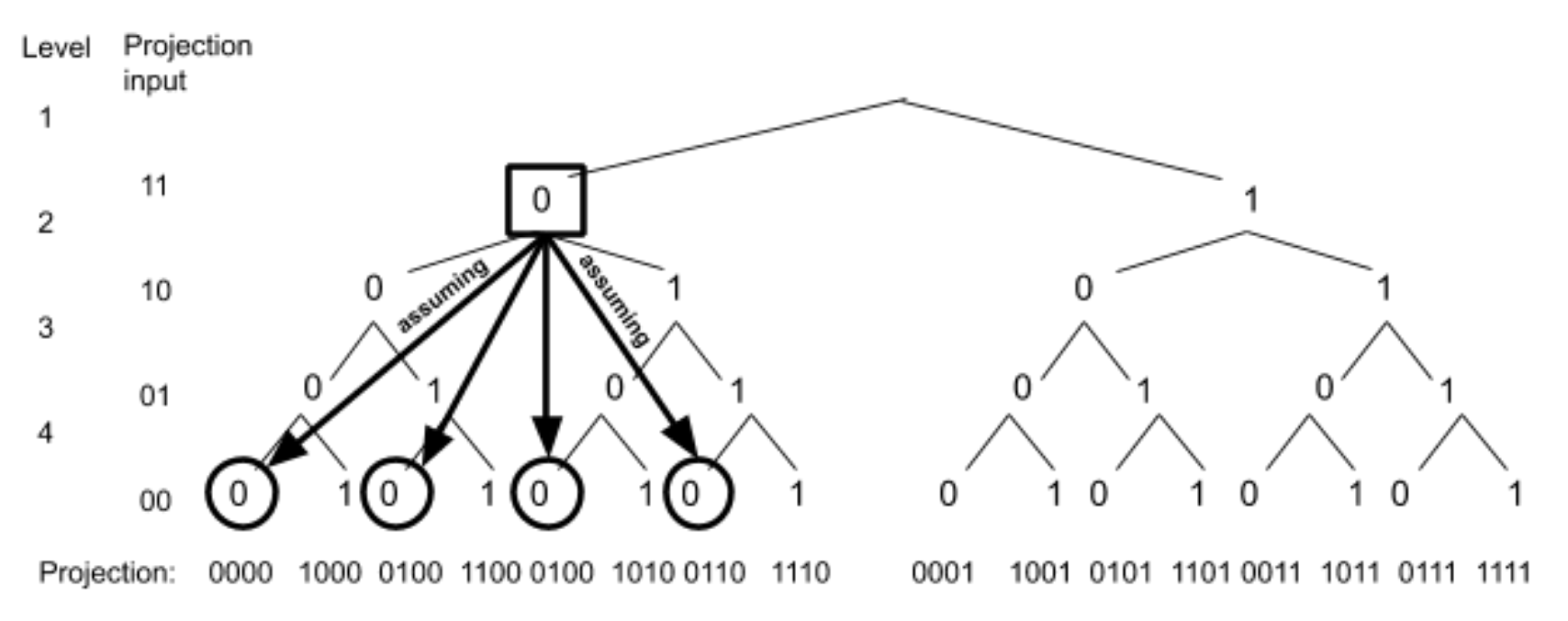}
\caption{\label{fig:assuming} Forced values. This search tree shows the coarse-graining from rule 2 to rule 4 at supercell size $N=2$. While calculating if we can continue after branching on $P(11)=0$, we find that the projection at level 4, $P(00)$ is forced to be 0. The algorithm thus records that $P(00)=0$ at this point, but then continues down the search tree as previously.}
\end{figure}

\begin{example} \label{ex:forced}
We illustrate the process of forcing on a coarse-graining from rule 2 to rule 4 with $N=2$; see Figure~\ref{fig:assuming}. In this example, we will assign variables in ``backwards'' order, beginning with $P(11)$ and ending with $P(00)$. We start by branching on $P(11)=0$. 
The only input for which $P(x_1), P(x_2), P(x_3)$ are all determined is $x=111111$. To find $res_1$ we apply rule 2 $N$ times and get the result of $f_A^N(x)=00$. However, $P(00)$ won't be determined until the 4th level of the tree, and we are only at level 1, so $P(00)$ is currently unassigned. We compute $res_2$ and get a result of 0. To ensure $res_1=res_2$, $res_1$ must also be 0, so we record $P(00)=0$. Then the algorithm continues by branching on the value of $P(10)$, then $P(01)$, and when it gets to the last level, it remembers that the value of $P(00)$ was already forced to $0$ and does not branch further; at that point it just checks whether the choices it has made satisfy (\ref{eq:CG}) for all inputs.
\end{example}

\subsection{Dynamic variable ordering}
Our third method to reduce computation time is to take further advantage of the forced values. Namely, when a value is forced, we can be more efficient by re-ordering the projection inputs dynamically (i.\,e., at the time). That is, instead of the levels of the tree being labeled in a static order such as 00, 01, 10, 11 (``forwards''), or 11, 10, 01, 00 (``backwards''), we can instead make the ordering dynamic as the algorithm proceeds. By considering a forced variable as soon as it is forced, the algorithm is able to make further inferences at the time, performing even more pruning and perhaps finding even more forced variables. This is analogous to unit propagation in SAT solvers.

\begin{figure}[!htbp]
\includegraphics[width=\columnwidth]{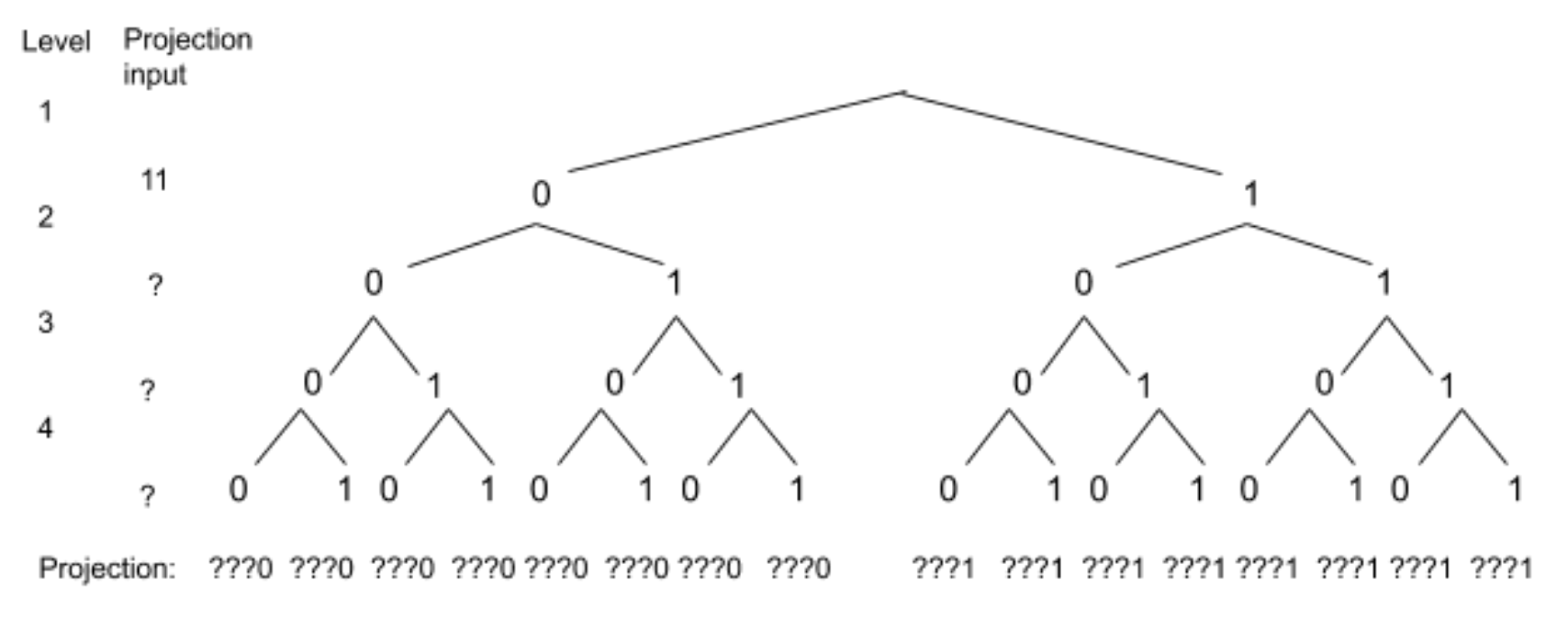}
\includegraphics[width=\columnwidth]{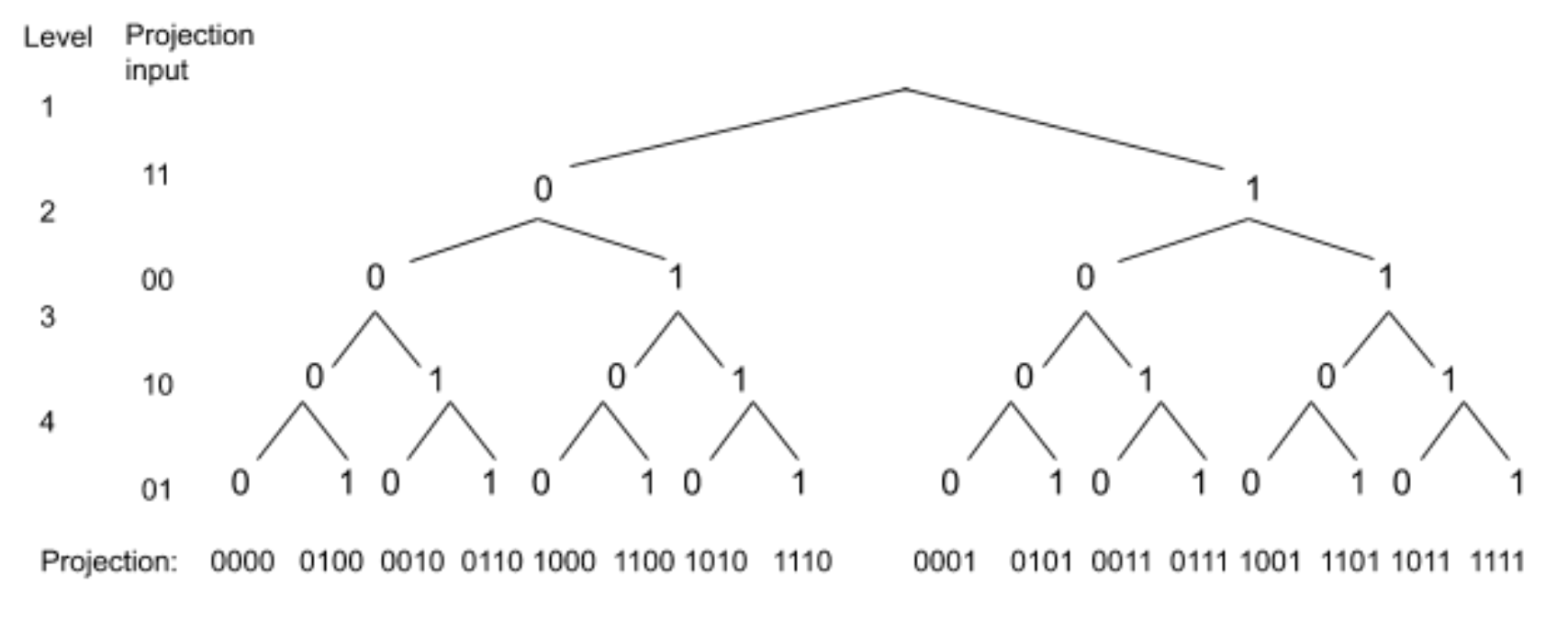}
\caption{\label{fig:dynamic} The dynamic ordering process. See Example~\ref{ex:dynamic}. (Above) In this example, the backwards algorithm is used, so it begins by branching on $P(11)$, while the ordering of the remaining variables is not yet decided (indicated by the ``?''s). (Below) After branching on $P(11)$, the value of $P(00)$ is forced, so level 2 is the assignments to 00. The process continues in this way until all the variables are assigned either by branching or being forced. }
\end{figure}

\begin{example}[Continuation of Example~\ref{ex:forced}] \label{ex:dynamic}
See Figure~\ref{fig:dynamic}. As in Example~\ref{ex:forced}, suppose we branch on $P(11)=0$ and discover that $P(00)=0$ is forced. Rather than waiting to see $P(00)$ at level 4 as ``originally planned'' (in the static, backwards order), the algorithm decides to make $P(00)$ the immediate next (second) level of the tree. The algorithm now continues as before. But while doing this, we see that $P(10)$ is also forced, so level 3 gets dynamically set to branch on $P(10)$. Dynamic ordering allows cuts to occur earlier in the search tree, significantly reducing the computation time. 
\end{example}

\subsection{Implementation variants}
Our dynamic ordering selects the next projection as one that is forced, if such a projection exists. We still have to explain what inputs $x$ to test $res_1 \stackrel{?}{=} res_2$ on (when there is a choice, which occurs frequently further along in the tree), and what projection to choose next when none are forced, including the first projection to branch on.

We implemented four specific strategies for these choices that we test experimentally:
\begin{itemize}
\item[$F_1$] In this strategy, the default ordering is ``forward,'' moving from $P(00\dotsb 0)$ to $P(11\dotsb 1)$, treating the input strings as binary representations of integers in the usual manner and going in increasing order of their integer values. In the first variant, after a forced variable is assigned, the next variable to branch on is simply the earliest unassigned input in the forward ordering.

\item[$F_2$] The default ordering is still ``forward,'' but now after a forced value is assigned, the next variable to branch on is the next one \emph{after} the forced variable in the forward ordering.

\item[$B_1$] Similar to strategy $F_1$, except the default ordering is ``backwards'', starting with $P(11 \dotsb 1)$ and ending with $P(00 \dotsb 0)$.

\item[$B_2$] Similar to strategy $F_2$, but with the default ordering being backwards.
\end{itemize}

For example, in strategy $F_1$, the algorithm begins by branching on $P(000)$. If $P(100)$ were then forced (and there were no variables forced immediately after), then the next projection to branch on would be $P(001)$, since $001$ is the first unassigned input in the forward ordering. In contrast, in the same scenario in $F_2$, the algorithm would branch next on $P(101)$, since $101$ is the next input after the forced input $100$.

\subsection{Rules 0 and 255}
Experimentally, we noticed that coarse-graining to rules 0 or 255 took a very long time, because pruning could not occur until very late in the search tree. Note that a nontrivial coarse-graining to (say) rule 0 is the same as a projection $P$ that is $0$ everywhere except possibly on some input $x$ that \emph{never occurs}  in the run of the CA $A$, that is, an input $x \in S_A^N$ that is not in the image of $f_A^N$. So for coarse-graining to rules 0 or 255, we instead search directly for such non-occuring inputs $x \in S_A^N$. When such an input is found, if coarse-graining to rule 0 we simply set $P(x)=1$, to ensure our projection is not the trivial all-0s projection, which recall we are excluding by fiat (for rule 255 we set $P(x)=0$ to avoid the all-1s projection).

If every possible projection input occurs at least once in the image of $f_A^N$, then we immediately conclude that $A$ cannot be non-trivially coarse-grained to either rule 0 or 255. Also, note that by this characterization, a rule $A$ can be non-trivially coarse-grained to rule 0 if and only if it can be coarse-grained to rule 255. (Note that for most $A$, this does \emph{not} follow from the fact that $\sigma_{01}(0)=255$, since that would only let us conclude that $A \to 0 \Leftrightarrow \sigma_{01}(A) \to 255$.)

\section{Results}
\subsection{Coarse grainings}
\begin{figure*}[!htbp]
\includegraphics[width=\textwidth]{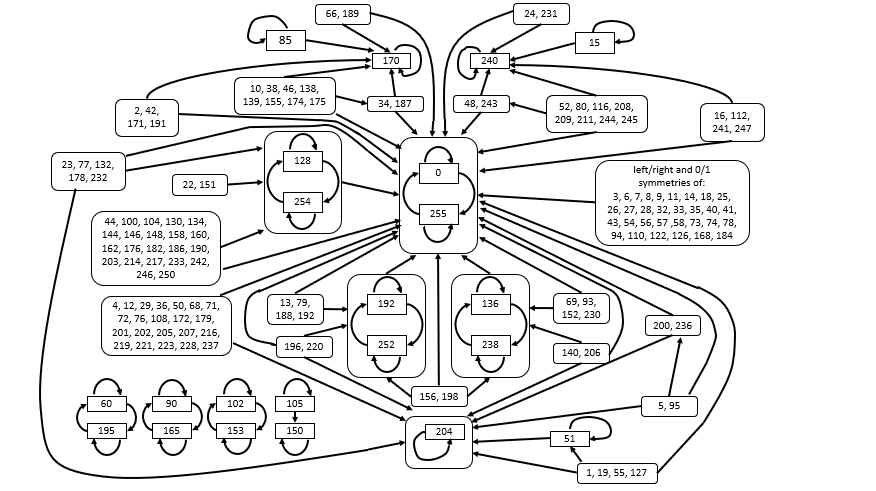}
\caption{\label{fig:diagram} Shows the coarse-graining transitions within the 256 elementary CA. Results from supercell size $N=2$ to $N=7$. An arrow indicates that the first rule may be coarse grained to the second rule for at least one choice of supercell size $N$ and nontrivial projection $P$.}
\end{figure*}

By implementing our improved method, we significantly sped up the search for coarse-graining. We used our new method(s) to exhaustively find all coarse-grainings between elementary CA with supercell size $N$ up to 7, on a commodity laptop. Israeli \& Goldenfeld reported results up to $N=4$, and briefly discussed further results they achieved with large amounts of time on a super-computer. Our Figure~\ref{fig:diagram} extends their results up to $N=7$.

The following coarse-grainings were discovered only at supercell sizes 5 or 6, but not smaller (CA are bracketed according to their symmetry classes):
\begin{itemize}
\item The following rules coarse-grain to rules 0 and 255: 
 [25, 61, 67, 103], [41, 97, 107, 121], [43, 113], [54, 147], [57, 99], [62, 118, 131, 145], [73, 109], 
 [94, 133], [104, 233], [110, 124, 137, 193], [122, 161], 
 [142, 212], 
\item The following coarse-grain to rule 204: 23, [36, 219], [50, 179], 77, [108, 201], [132, 222], 178, 232
\item $[1,127], [19, 55] \to 51$
\item $[22,151], [104, 233] \to [128, 254]$
\item $[66, 189] \to 170$, and the symmetric $[24,231] \to 240$
\end{itemize}
After $N=5,6$, despite exhaustively and conclusively searching at $N=7$, we did not discover any new pairs $(A,B)$ such that $A$ coarse-grained to $B$ at supercell size $7$ but not smaller. 

The main diagram of results in \cite{IG1, IG2} has 54 elementary CA that have no nontrivial coarse-grainings at all up to supercell size 4; using a super-computer to explore larger supercell sizes, Israeli \& Goldenfeld then report that there were only 16 elementary CA for which they couldn't find a nontrivial coarse-graining. On a laptop, we find only 20 elementary CA that do not have coarse-grainings up to supercell size $N=7$: the 16 previously reported, along with the two symmetry classes [37,91] and [164, 218]. The non-trivial coarse-grainings for these four are presumably present at larger supercell sizes.

\subsection{Comparison of algorithms}
We compared the four implementation variants $F_1, F_2, B_1, B_2$ of our method discussed above with the brute force method of \cite{IG1, IG2}. 
The brute force approach does not include any of the new ideas developed in this paper, such as the search tree or pruning branches. The brute force method does, however, include the reduction due to symmetries. 

\begin{figure}[!htbp]
\includegraphics[width=\columnwidth]{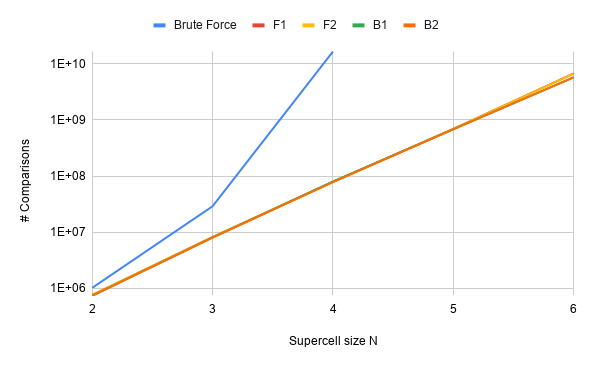}
\caption{\label{fig:comparison} Number of comparisons $res_1=res_2$ used by each of the five methods---brute force, $F_1, F_2, B_1, B_2$---in finding coarse-grainings up to supercell size $N=6$. (The plots for $F_1, B_1$ aren't visible simply because they're so well-overlapped by those for $F_2, B_2$ at this scale.) Note the $y$-axis is scaled logarithmically, so all four of our new approaches appear to exhibit singly exponential scaling (approximately $\propto 10^N$), in comparison to the $2^{2^N}$ scaling of the brute force algorithm. The counts for the brute force algorithm are not reported for $N > 4$ as the algorithm took too long to run.
}
\end{figure}

To measure efficiency we will display the number of comparisons between $res_1$ and $res_2$; these are displayed in Figure~\ref{fig:comparison}; since the four graphs for $F_1, F_2, B_1, B_2$ were so similar, we also report the numerical values for the five approaches in Table~\ref{table:comparison}.

\begin{table}[!htbp]
\begin{tabular}{r|ccccc}
$N$ & Brute Force & $F_1$ & $F_2$ & $B_1$ & $B_2$ \\ \hline
2 & 1,013K & 755K & 755K & 731K & 731K \\
3 & 28,942K & 8,169K & 8,165K & 7,990K & 8002K \\
4 & 16,687M & 78,964K & 78,830K & 79,954K & 77,999K \\
5 & * & 694M & 694M & 685M & 686M \\
6 & * & 6,778M & 6,722M & 5,777M & 5,798M
\end{tabular}

\caption{\label{table:comparison} The number of comparisons of $res_1$ and $res_2$ for five methods (brute force, $F_1$, $F_2$, $B_1$, $B_2$). The * indicates values left uncomputed because the computations took too long.}
\end{table}

    As expected, the brute force approach was significantly less efficient than the other 4 selection methods. In Figure~\ref{fig:comparison} all four of our new approaches appear to exhibit singly exponential $\exp(cN)$ scaling, in comparison to the $2^{2^N}$ scaling of the brute force algorithm. As we progress, pruning branches apparently has greater effect in reducing the number of comparisons.
    
Among $F_1$, $F_2$, $B_1$, and $B_2$, we see that despite $B_1$ being the best approach overall up to $N=6$ there is only about a 1-2\% difference in the number of comparisons for $N=4$ and 5. When we move on to $N=6$, however, the difference increases to about 15\%. On examination, we found that this greater difference was caused by a single case, the coarse graining from rule 162 to rule 170. If we eliminate this case, $N=6$ would also have a miniscule 1-2\% difference between our four variants, similar to that observed at smaller values of $N$.

The coarse-graining from rule 162 to rule 170 at $N=6$ is an interesting example because of the immense difference between computations in the forward and backward direction. When applying $B_1$, there were precisely 500K comparisons. Meanwhile, for $F_1$, there were 916M comparisons, a factor of approximately 1800. 
In this specific example, the backward direction yields a drastic improvement over the forward direction, but this does not hold true for all other cases.

In the coarse-grainings from rule 154 to rule 170 with $N=4$, we discovered that it was the opposite situation in which the forward direction was far more efficient than the backward direction. When using the backward approach, 66,040 total comparisons were needed to complete. Meanwhile, when we tried the forward approach, there were only 7,128 comparisons, a factor of just under 10. 

We also find examples where there is a significant difference between selection methods 1 and 2. In coarse-graining rule 29 to rule 51 with $N=7$, method $B_1$ used 1,353K comparisons, while $B_2$ used only 241K, a nearly six-fold difference. 

More importantly, we discovered that by altering the initial input projection, we could reduce the number of comparisons by an extremely large amount. Although it was a coincidence, we found that when choosing to first branch on $P(1010101)$, only 60 comparisons were required to complete (just 60, not 60K!). Even if we do not expect such drastic improvements in all cases, the magnitude of this drop was surprising, and suggests that heuristics for better selection of projection inputs to branch on could yield quite significant further improvements.

As is to be expected, no one ordering is always the best. Thus, to provide optimal efficiency we could combine all four of these. Each of the coarse-grainings are more effective with different types of dynamic orderings due to their unique binary patterns, so to further reduce the computations, we still need to discover how to choose a selection method that will be most compatible with each of the rules. We now discuss a few patterns we noticed that might be useful in this endeavor.

Using these new approaches, we observed four common patterns in the cuts. The first in which both the 0 and 1 branches immediately prune off at level 1, the lowest level. The second pattern tapers from higher to lower levels in the 0 branch but cuts immediately at level 1 in the 1 branch. 
The third pattern is the opposite, in which 0 immediately cuts at level 1 but the 1 branch tapers from lower to higher levels. 
The last pattern we saw frequently was a combination of the second and third, in which the 0 branch tapers from higher to lower levels while the 1 branch tapers from lower to higher levels. 

While most coarse grainings followed these four general trends with slight deviations, there were some outliers. The coarse-graining from rule 162 to rule 170 in the forward ordering was unique among all those we explored. It had a seemingly highly irregular pattern of cuts, with most cuts occurring only very late in the tree, having little effect in reducing the runtime. However, when we tried reversing the order, starting at 11, the shape transformed into the fourth pattern mentioned above and the computation finished fairly quickly. 

Overall, we saw that when there were cases in which the coarse-graining did not follow one of the preceding four patterns, they far took longer to compute than others. However, as seen in the example above, changing the direction can sometimes help fix this issue.

\section{Conclusion}
We developed a much more efficient algorithm for finding coarse-grainings between cellular automata, using a backtracking search, with propagation of forced values and a dynamic variable ordering. Experimentally, our method appears to have singly-exponential time scaling, compared to the previous brute-force method whose runtime scaled as $2^{2^N}$ \cite{IG1, IG2}. Using our method we could examine exhaustively all possible coarse-grainings of elementary cellular automata up to supercell size 7, extending the previous results up that only went up to supercell size 4 \cite{IG1, IG2}. We found 26 new symmetry classes of coarse-grainings (56 coarse grainings total). 

To explore further improvements to our algorithm, we examined several different variable orderings, and found that each had different advantages. We examined specific cases that suggest that heuristics to select which variable to branch on next could have quite drastic effects on the runtime. Other possible improvements are suggested by analogy with the SAT literature, such as an analogue of clause learning, or to try to incorporate SAT solvers directly but dynamically (rather than an all-at-once reduction, see Remark~\ref{rmk:SAT}).

Interesting open questions include the original questions raised in \cite{IG1, IG2}---such as whether the remaining 16 elementary CA can be coarse-grained at all---as well as new questions we highlight, such as classifying the complexity of the coarse-graining problem. The experimentally observed singly-exponential scaling of our algorithm raises the possibility that the problem might in fact be in the complexity class $\mathsf{EXP}$, making the question of whether it is in $\mathsf{EXP}$ or is $\mathsf{NEXP}$-complete even more salient. There also remains the question finding mathematical methods to prove when there does not exist a coarse graining between two given (elementary) cellular automata---\emph{regardless} of supercell size---which would enable us to complete the picture of coarse-grainings between elementary CA.

\begin{acknowledgments}
We would like to thank \href{https://www.athenabywistem.org/}{ATHENA} by \href{https://womeninstem.org/}{WiSTEM}, the organization that matched the authors together, eventually leading to this project. The authors were partially funded by NSF grant DMS-1829826 (formerly DMS-1622390).
\end{acknowledgments}

\bibliography{CA}

\end{document}